# Characterization and Characteristics of mechanochemically synthesized amorphous fast ionic conductor 50 SISOMO ($50AgI-25Ag_2O-25MoO_3$)

**Saurabh Dayal, and K. Shahi** *

*Department of Physics & Material Science Program, Indian Institute of Technology, Kanpur, India-208016*

**Abstract**

Mechanochemically synthesized amorphous 50SISOMO [$50AgI-25Ag_2O-25MoO_3$] fast ionic conductor shows high ionic conductivity of ~ $6x10^{-3}\ \Omega^{-1}\ cm^{-1}$ at room temperature. The highest ionic conductivity is achieved for 36 h milled sample, which is more than three orders of magnitude higher than that of crystalline AgI at room temperature. The samples are thermally stable at least up to ~70 °C. Thermoelectric power studies on 50 SISOMO amorphous fast ionic conductors (a-SIC) have been carried out in the temperature range 300-330K. Thermoelectric power (S) is found to vary linearly with the inverse of the absolute temperature, and can be expressed by the equation $-S = [(0.19 \times 10^3/T) + 0.25]$ mV/K. The heat of transport ($q^*$) of $Ag^+$ ion i.e. 0.19 eV is nearly equal to the activation energy ($E_\sigma$) i.e. 0.20 eV of $Ag^+$ ion migration calculated from the conductivity plots indicating that the material has an average structure. This is also in consonance with earlier theories on heats of transport of ions in ionic solids.

## 1. Introduction

Ionic Conduction in solids has been a subject of interest for as early as the beginning of 19th century. In recent years, much attention is focused on fast ionic conducting glasses because of their exceptionally high ionic conductivity, potential and application in batteries and other electrochemical devices. Especially, AgI doped silver oxysalt glasses, such as $AgI-Ag_2O–M_pO_q$ (M= Mo, V, Bo etc.) systems, are much investigated, since these system contain AgI, which is a representative of fast ionic conductor, and possess high ionic conductivity near room temperature. AgI doped fast ionic conducting glasses have several common features. The ionic conductivity increases with increasing AgI concentration and reaches $10^{-2}$ S $cm^{-1}$ for these glasses having high AgI concentration. Ionic motion is a very interesting and challenging problem in condensed matter science. More recently, high energy ball-milling at room temperature has been successfully used to form amorphous fast ionic conductors FICs [1-4]. The mechanochemically-milled (MM) glasses generally have a higher conductivity, lower glass transition ($T_g$) and crystallization ($T_c$) temperatures.

* Corresponding author. E-mail address: kshahi@iitk.ac.in (K. Shahi).

Importantly, the dc conductivity ($\sigma_{dc}$) vs. inverse temperature behaviour of MM glasses is found to exhibit two Arrhenius regions, a low temperature region with lower activation energy and a high temperature region with higher activation energy.

The frequency dependence of the real part of the complex conductivity

$$\sigma^* = \sigma^{/}(\omega) + j\, \sigma^{//}(\omega) \quad (1)$$

is useful in understanding the mechanism of ion dynamics and various processes [8]. Within a given solid, temperature and excitation frequencies can trigger a whole range of ion dynamic phenomenon. Furthermore, the frequency dependent conductivity of several of these complex systems exhibits universal behavior.

## 2. Experimental Details

The mixture was taken in agate pot of size 250 ml with 4–6 number of agate balls of diameter ~ 20 mm. Milling was performed in acetone medium in a planetary ball-mill (Fritsch, puleverisette−6) operating at 400 rpm. The amount of acetone was 10–15 ml for each 8 g of mixture batches and was monitored after each 3 h to keep it constant. The samples are examined by X-ray diffraction for their non-crystalline nature. The impedance (Z) and phase angle (θ) data are measured using HP-4192A impedance analyzer interfaced with a PC using Lab view program in the frequency range 10 Hz- 10 MHz in the temperature range of 100-340 K using a cryostat system. These all measurements are done on highly pressed pallets having the pressed silver powder electrodes. The real part of the complex conductivity also known as ac conductivity is calculated with the help of the equation

$$\sigma' = \left(\frac{\cos\theta}{Z}\right)\frac{L}{A} \quad (2)$$

where L is the thickness and A is the area of cross section of the sample.

The thermopower studies were carried with the help of small heater attached with the one electrode of the sample holder prepared. The voltage at the internal heater was supplied by the ac power supply (0-12V). The voltage across the two thermocouple attached to the two electrode was measured with the help of Digital Multimeter (DMM Philips).

## 3. Results and discussion

### *3.1 Impedance analysis and dc conductivity*

Fig 1 shows the variation of real (Z cosθ ) vs. imaginary (Z sinθ) parts of impedance for 50 SISOMO. At low temperature, the impedance data is depressed semicircle but as the temperature increases the plots are no more semicircles.

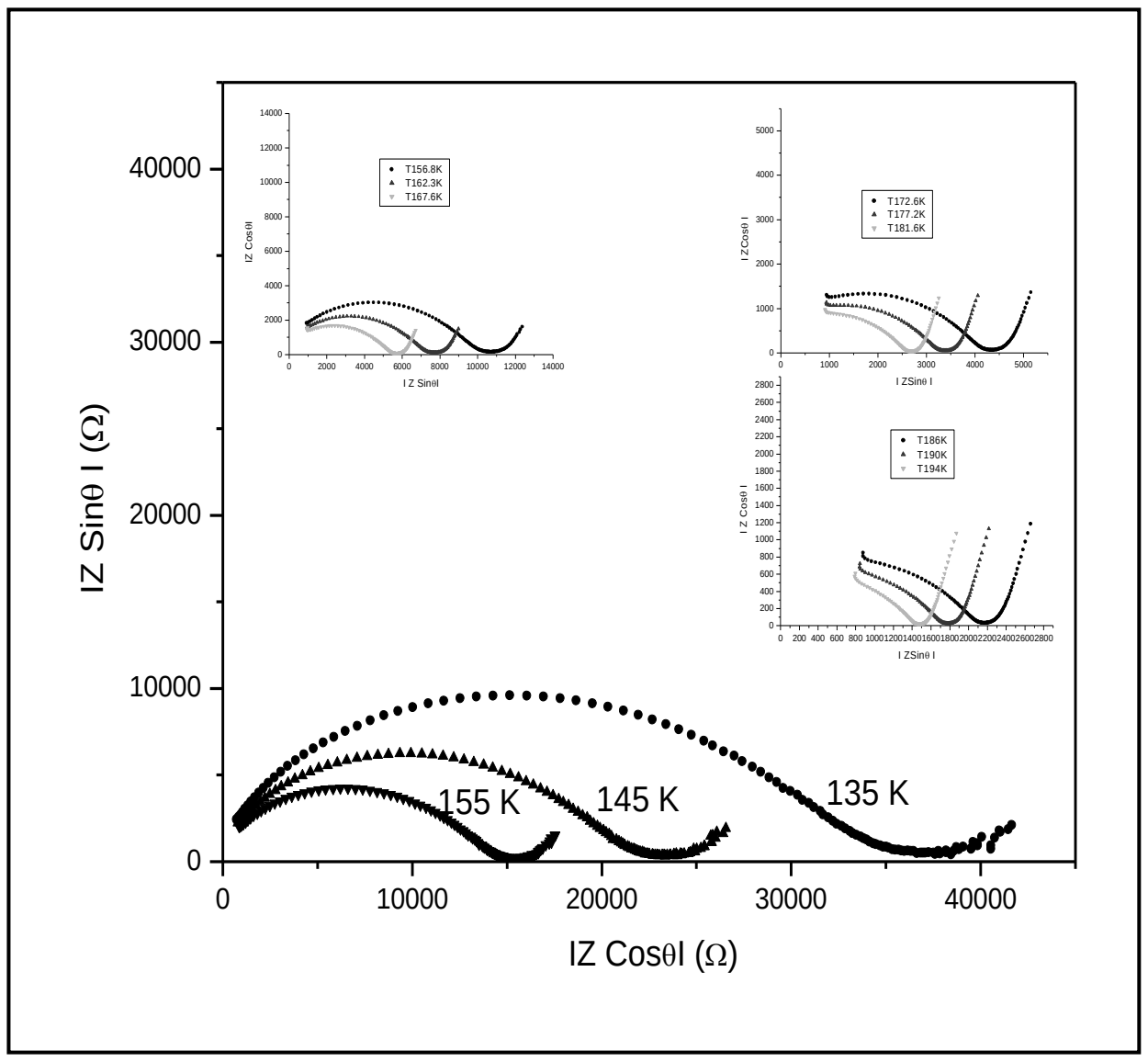

Fig 1: Impedance plots for the mechanochemically synthesized 50 SISOMO amorphous fast ionic conductors at different temperatures. Insets of the plot shows the temperatures for which complex spectra are not accessible

Fig. 2 shows the variation of dc conductivity as function of inverse temperature for 50 SISOMO. It can be concluded from the plot that it has two activation energies in two different temperature ranges and hence two Arrhenius regions. The lower activation energy is 0.13 eV in lower temperature range (100-220K) and the higher one is 0.20 eV in higher temperature region (221-300K).It is concluded here that these two distinct Arrhenius regions in the $\sigma_{dc}$(T) are associated with two different sets of $Ag^+$ ions.

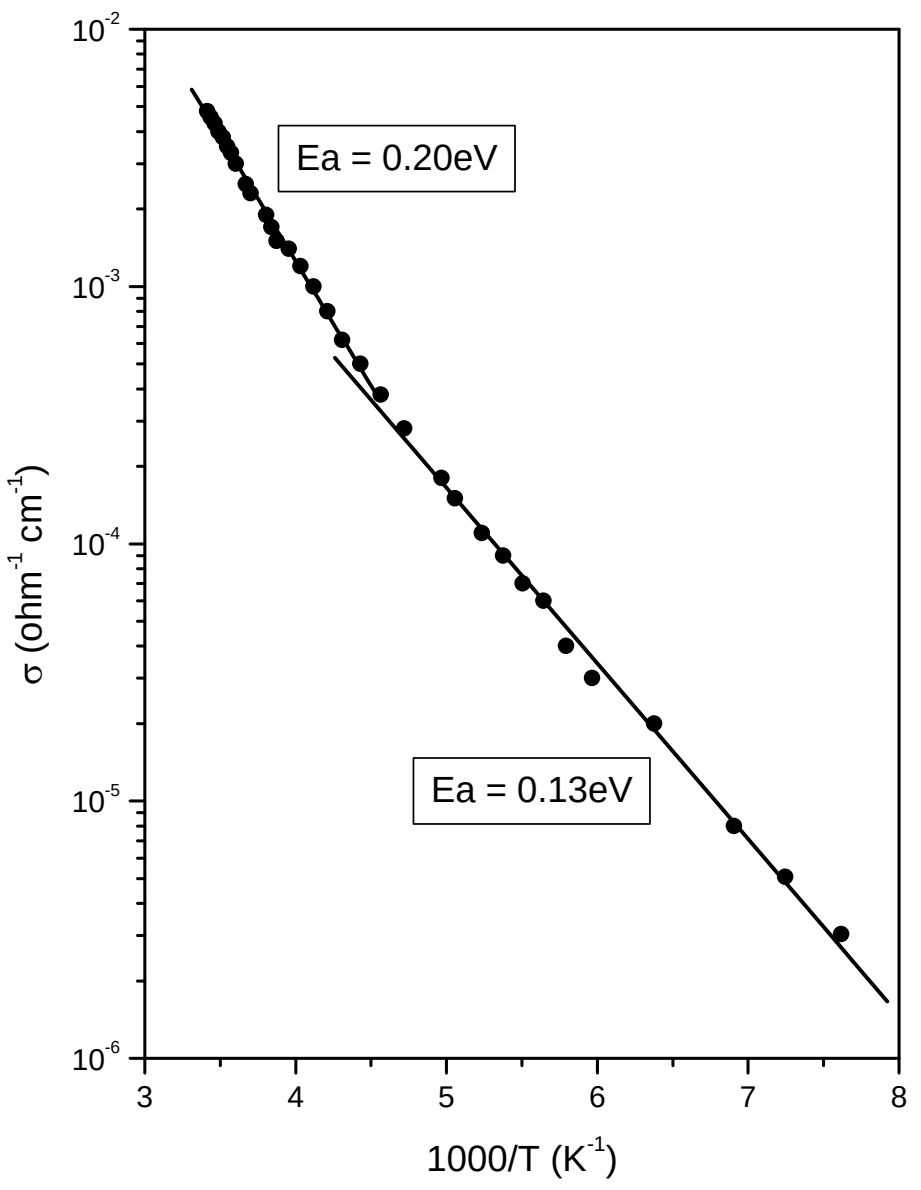

Fig 2: The dc conductivity ($\sigma_{dc}$) vs. 1000/T plot for 50SISOMO amorphous fast ionic conductors, showing two different Arrhenius regions with different activation energies.

*3.2 AC electrical conductivity*

*3.2. (a) Conductivity isotherms*

The frequency dependence of the real part of the complex conductivity $\sigma^{/}(\omega)$ for 50 SISOMO is shown in fig.3 at several temperatures. At low frequencies random diffusion of the ionic charge carriers via activated hopping gives rise to the frequency independent nature [6-9]. But as frequency increases the ac conductivity shows a dispersion, which shifts to higher frequencies as temperature increases. Many of the compositions of SISOMO [10] are following the universal dynamic response. This behaviour of ac conductivity of 50 SISOMO is also well accounted by the universal dynamic response (UDR) [5], in which the ac conductivity is given by the following power law

$$\sigma(\omega) = \sigma_{dc} + A\,\omega^{n} \qquad (3)$$

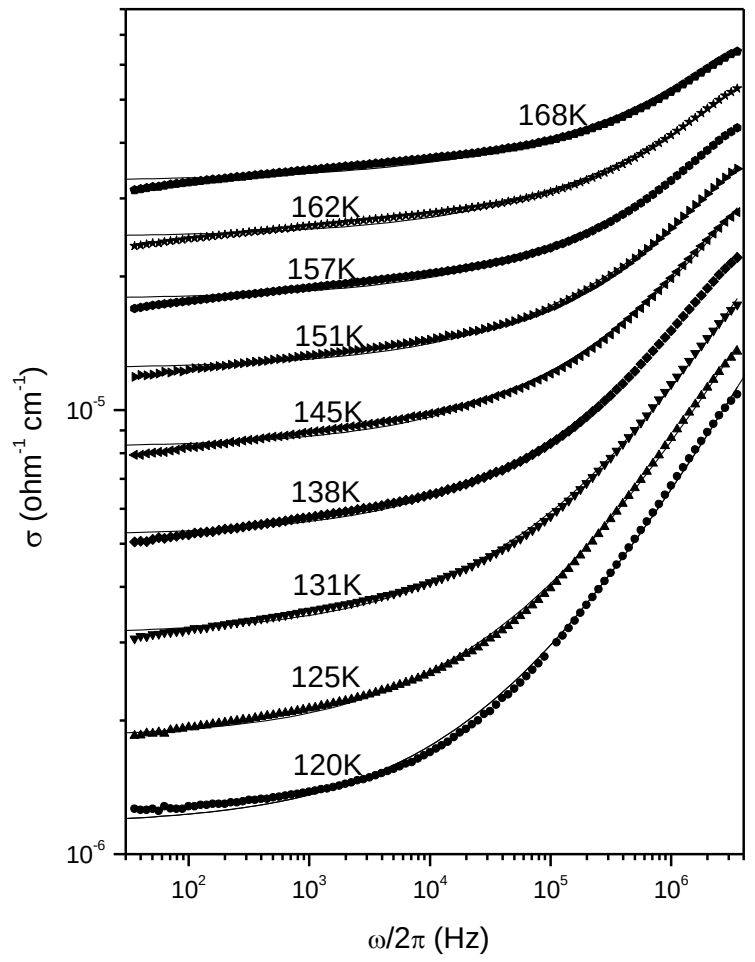

Fig 3: The ac conductivity spectra for 50 SISOMO at different temperatures. The solid lines are best-fit curves to the experimental data according to UDR. $\sigma(\omega) = \sigma_{dc} + A\,\omega^{n}$

*3.2. (b) Isochronal Curve*

The ac conductivity vs. 1000/T plot at constant frequencies is shown in fig 4. It is clear that at high temperature the conductivity is following straight-line nature but in low temperature region there is some dispersion. It is clear that the ac conductivity exhibits limited regions of linear Arrhenius behavior. It is important here that at low temperature end the conductivity is hardly changes with temperature but shows strong dependence on frequency [8]. This phenomenon is currently under active discussion and is often described in terms of nearly constant loss and superlinear power law [8]. The results shown in Fig 4 also indicate the possible existence of superlinear power law in these mechanochemically synthesized 50 SISOMO amorphous fast ionic conductors.

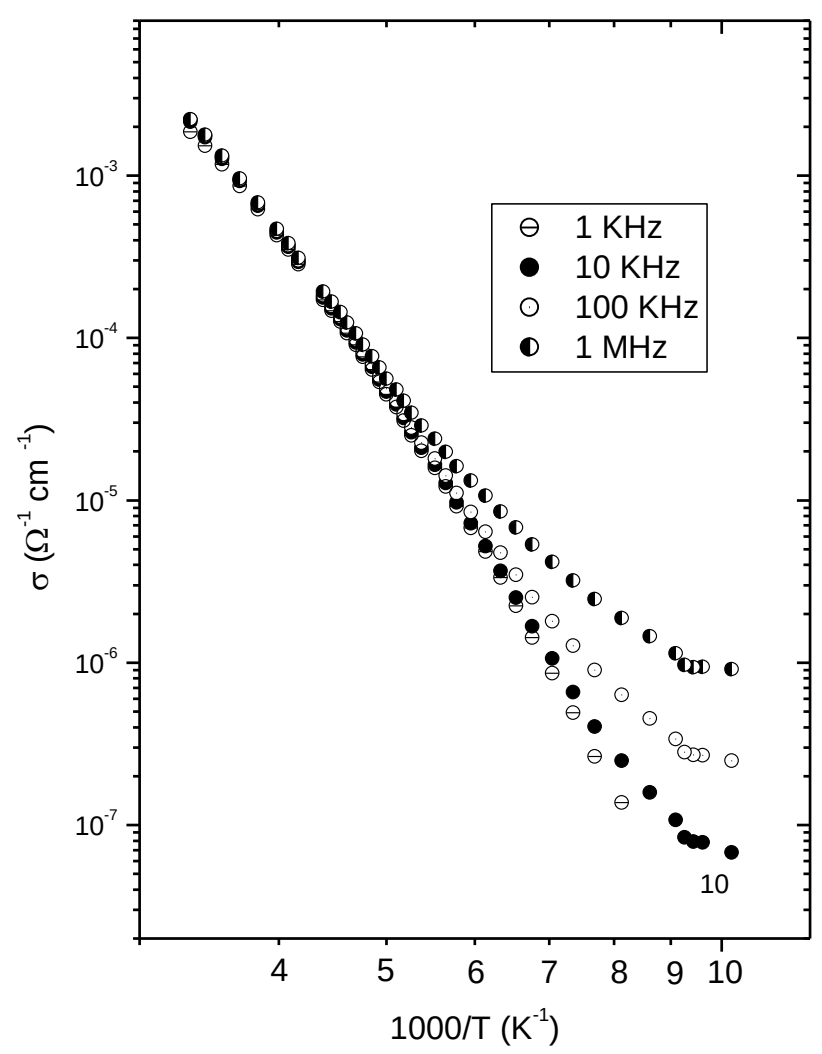

Fig 4: ac conductivity of 50 SISOMO amorphous fast ionic conductors at various frequencies

### 3.3 Thermopower

The variation of thermoelectric voltage with temperature difference for 50 SISOMO amorphous fast ionic conductors is shown in Fig.5. The thermoelectric power (S) is the measure of the thermoemf across the sample pallet with a constant temperature difference. The thermoemf - temperature difference plot is a straight line and the slope is thermopower (S) = 0.35 mV/K.

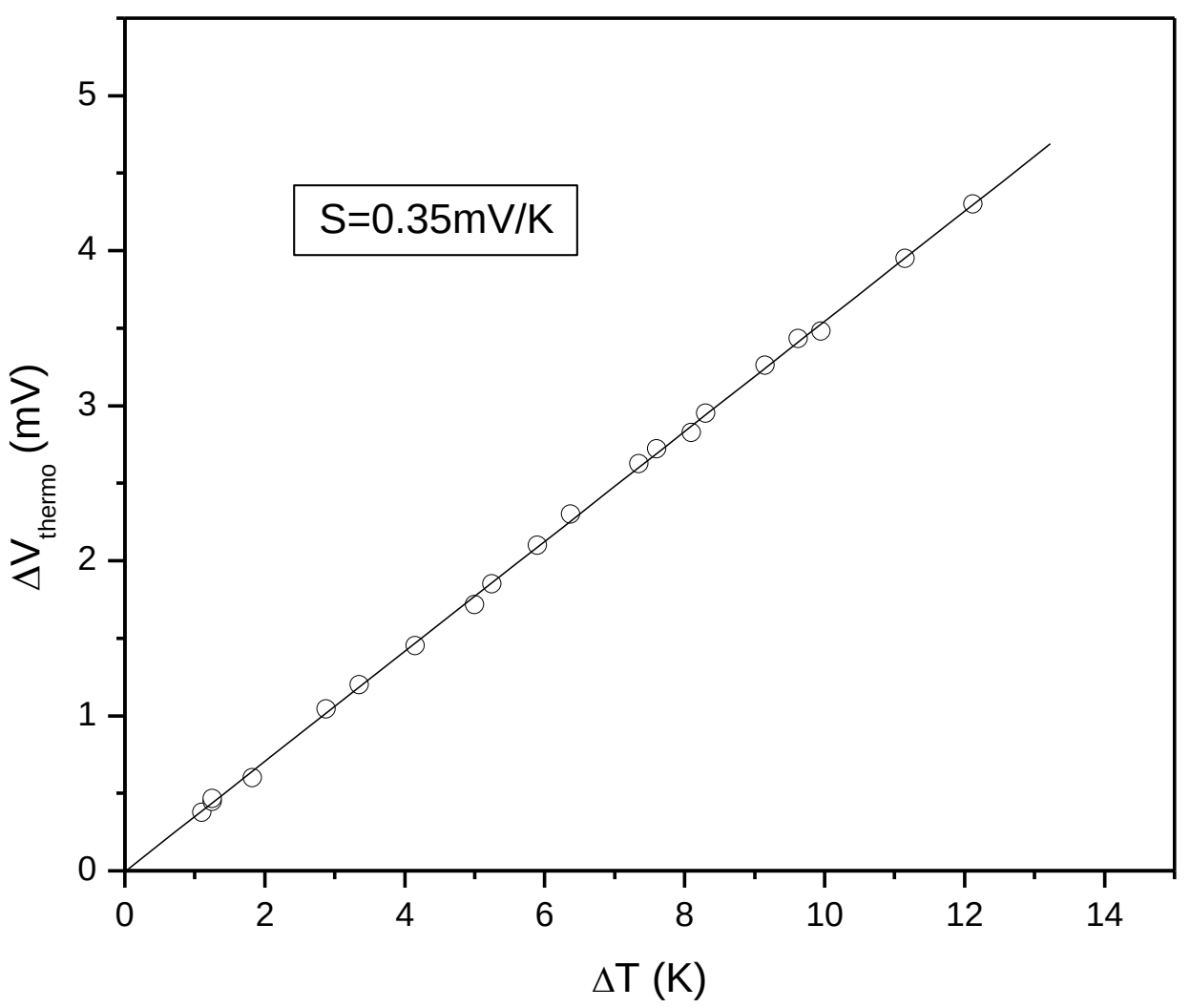

Fig 5: Variation of thermo emf with temperature difference for 50 SISOMO amorphous fast ionic conductor.

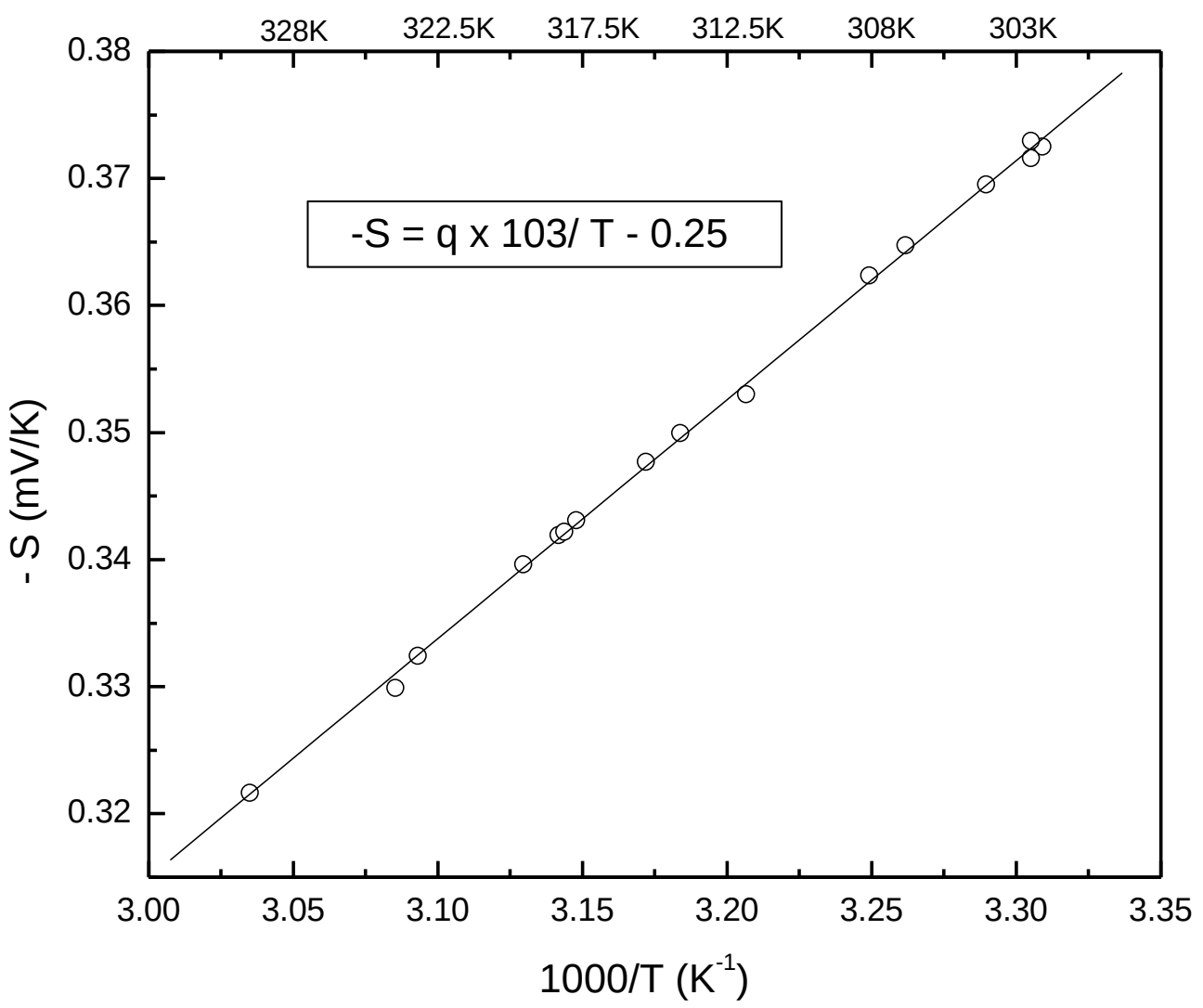

Fig 6: Variation of thermopower with inverse of temperature for 50 SISOMO

Fig 6 shows the thermopower (-S) - inverse temperature plot. This plot is also straight line and the slope of the plot is actually the heat of transport (q) of the sample i.e.0.19. It is very close to the activation energy (0.20 eV) of the sample in this temperature range (300-330 K). It can be concluded here that heat of transport of 50 SISOMO amorphous fast ionic conductors is same to the activation energy in the temperature range. It is indicating that the material has an average structure. This is in accordance to the well known theory of Rice and Roth. The relation governing the thermopower with temperature inverse can be written as

$$-S = q \times 10^3 / T - 0.25. \tag{4}$$

## 4. Conclusions

The 50 SISOMO amorphous fast ionic conductors have the ionic conductivity of ~6 x $10^{-3}\ \Omega^{-1}\ cm^{-1}$ at 300 K. The impedance plot shows depressed semicircle at low temperature. The conductivity calculated with the help of impedance plot is found to be very close to the value which is calculated with the help of the power law fitting. While the well known UDR power law $\sigma(\omega) = \sigma_{dc} + A\ \omega^n$ does reproduce the frequency dependent conductivity. For 50 SISOMO the activation energy in the low temperature range (100-220 K) is found to be 0.13 eV while it is somewhat higher 0.20 eV at high temperatures (221 -340K).These two Arrhenius region are essentially due to the existence of two different bonding states of $Ag^+$ ions.

Thermopower of the sample is calculated to -0.35mV/K.The heat of transport (q) of the sample is 0.19eV. It is very close to the activation energy (0.20 eV) of the sample in this temperature range. The equation relating thermopower with heat of transport can be written as $-S = q \times 10^3 / T - 0.25$. It is indicating that the material has an average structure. This is also in consonance with earlier theories (Rice and Roth) on heats of transport of ions in ionic solids.

## References

1. T. Minami, H. Nambu, M.Tunaka, J. Electrochem. Soc. 60 (5-6) (1977) 283
2. K. Shahi Phys. Status Solidi, A Appl. Res 41 (1977) 11
3. A. Dalvi, K.Shahi, Solid State Ionics 148 (2002) 431-436
4. A. Hayashi, H. Shigenoi, T. Minami, M. Tatsumisago, Electrochem. Commun. 9 (2003) 111-114
5. A. Dalvi, K.Shahi, J. Phys. Chem. Solids 64 (2003) 1-7
6. A. Dalvi, K. Shahi, Solid State Ionics 159 (2003) 369-379
7. A.K. Jonscher, Nature 267 (1977) 673.
8. J.P.Tiwari, K.Shahi, Solid State Ionics (Article in press)
9. M.C.R. Shastry, K.J.Rao, Solid State Ionics 44 (1991) 187-198
10. Saurabh Dayal, K.Shahi, work in progress.